\documentclass[12pt,preprint]{aastex}

\newcommand{\beq}{\begin{equation}}
\newcommand{\eeq}{\end{equation}}

\newcommand{\etal}{et~al.~}

\newcommand{\ms}{\mbox{m s$^{-1}~$}}

\newcommand{\msun}{M$_{\odot}~$}

\newcommand{\mjup}{M$_{\rm JUP}~$}

\newcommand{\mearth}{M$_{\rm EARTH}$}
\newcommand{\mnep}{M$_{\rm NEPTUNE}$}

\newcommand{\msini}{$M \sin {\it i}~$}
\newcommand{\asini}{$A \sin {\it i}~$}

\newcommand{\rc}{$\rho^{1}$\,Cancri~}
\newcommand{\aaph}{A\&A}

\shorttitle{Low Mass Planet in the \rc system} 
\shortauthors{McArthur et al.}

\begin{document}


\title{Detection of a NEPTUNE-mass planet in the \rc
  system using the Hobby-Eberly Telescope}


\author{Barbara E. McArthur\altaffilmark{1}, Michael Endl\altaffilmark{1}, 
 William D. Cochran\altaffilmark{1}, and G. Fritz Benedict\altaffilmark{1}}

\author{Debra A. Fischer\altaffilmark{2},Geoffrey W.  Marcy\altaffilmark{2,3},  
 and R. Paul Butler\altaffilmark{4}}

\author{Dominique Naef\altaffilmark{5,6}, Michel Mayor\altaffilmark{5}, 
Diedre Queloz\altaffilmark{5}, and Stephane Udry\altaffilmark{5}}

\and

\author{Thomas E. Harrison\altaffilmark{7}}


\altaffiltext{1}{McDonald Observatory, University of Texas,
    Austin, TX 78712}

\altaffiltext{2}{ Department of Astronomy, University of California, 601 
Campbell Hall, Berkeley, CA USA 94720}

\altaffiltext{3}{ Department of Physics and Astronomy, San Francisco
State University, 1600 Holloway Avenue, San Francisco, CA USA 94132}

\altaffiltext{4}{ Department of Terrestrial Magnetism, Carnegie Institution
of Washington, 5241 Broad Branch Road NW, Washington DC, USA 20015-1305}

\altaffiltext{5}{Observatoire de Geneve, 51 Ch. des Maillettes, 1290 Sauverny,
 Switzerland } 

\altaffiltext{6}{ESO, Alonso de Cordova 3107, Casilla 19001, Santiago 19, Chile}

\altaffiltext{7}{Department of Astronomy, New Mexico State University, 
1320 Frenger Mall, Las Cruces, New Mexico 88003}



\begin{abstract}
We report the  detection of the lowest mass 
extra-solar planet yet found
around a Sun-like star - a planet  with an \msini of only
14.21 $\pm$ 2.91 Earth masses in an  
extremely short period orbit (P=2.808 days) around  \rc,
a planetary system which already has
three known  planets. Velocities taken from late 2003-2004 at
McDonald Observatory with the Hobby-Eberly Telescope (HET) revealed 
this inner planet at 0.04 AU.  We estimate an inclination of the outer
planet \rc d, based upon
{\it Hubble Space Telescope~} Fine Guidance Sensor (FGS) measurements
which suggests  an inner  planet of only 17.7  $\pm$ 5.57 Earth
masses, if coplanarity is assumed for the system.

\end{abstract}


%% subject header.  Objects should be in the appropriate "individual"
%% headers (e.g. quasars: individual, stars: individual, etc.) with the
%% additional provision that the total number of headers, including each
%% individual object, not exceed six.  The \objectname{} macro, and its
%% alias \object{}, is used to mark each object.  The macro takes the object
%% name as its primary argument.  This name will appear in the paper
%% and serve as the link's anchor in the electronic edition if the name
%% is recognized by the data centers.  The macro also takes an optional
%% argument in parentheses in cases where the data center identification
%% differs from what is to be printed in the paper.



\section{Introduction}

Of the 123 planets found so far around 108 main-sequence stars, there are 13
multiple planet systems. The fourth extrasolar planet discovered was 
the 14.65 day period
planet around \rc ($=$ rho Cancri A $=$ 55 Cancri 1$=$ HD 75732 $=$ HIP 43587
$=$ HR3522) \citep{But97}. More recently, the orbits of an additional outer planet and possible 
intermediate planet have been reported \citep{Mar02}.  In this Letter, 
we announce the discovery of a fourth inner
planet in the \rc system with radial velocity semi-amplitude (K) of 
6.7 \ms, an orbital period of 2.808 days, \msini of 0.04  Jupiter 
masses, and a probable actual mass of 0.056 Jupiter
masses (17.7 Earth masses).

\rc  appears to be a super metal-rich (G8V, V = 5.95, [Fe/H] = +0.27),
but otherwise normal main sequence star with a V=13 visual 
binary companion.  There is no
detectable circumstellar disk around \rc \citep{Jay02, Sch01}.  The 
metallicity  is significantly higher 
than the mean of nearby field stars, but is in accord with
the trend of close-in giant planets being found preferentially 
around metal-rich solar-type and cooler
stars \citep{Tay96,Gon98,Fel01,Cay01,Law03,Hei03}.  
Mass estimates for \rc range from 0.88 - 1.08 \msun \citep{Law03,Fuh98,For99,San03}.
We adopt a mass
of  0.95 \msun $\pm$ 0.08 (Allende-Prieto 2004, personal communication).

\section{Observations and Results}

%% In a manner similar to \objectname authors can provide links to dataset
%% hosted at participating data centers via the \dataset{} command.  The
%% second curly bracket argument is printed in the text while the first
%% parentheses argument serves as the valid data set identifier.  Large
%% lists of data set are best provided in a table (see Table 3 for an example).
%% Valid data set identifiers should be obtained from the data center that
%% is currently hosting the data.

\subsection{{\it HST} Astrometry}
This Letter made use of  reanalyzed public domain data (McArthur \etal
2005,in  preparation) obtained from the Hubble Space 
Telescope ({\it HST}) \citep{McG02,McG03} 
to provide an estimate
of the inclination of the long period planet, \rc  d.  For the 
parameters critical in
determining the mass of \rc d, we find a parallax,
$\pi_{abs}$ = 79.78 $\pm$ 0.3 mas, a
perturbation size, $\alpha$= 1.94 $\pm$ 0.4 mas, and an 
inclination, i = 53$\arcdeg$ $\pm$ 6.8$\arcdeg$  
This inclination
was derived from a small arc of orbit coverage in the limited {\it HST} 
data set.  
In any case, an inclination of less than 20$\arcdeg$ can not be considered  
because of the very large perturbation that could not have been missed, 
even by the small existing sample of {\it HST} astrometric data.

\subsection{Radial Velocity}

We used the Hobby-Eberly Telescope (HET) to make many, high precision
radial velocity observations of \rc  over a short time period in 
preparation for
anticipated  additional {\it HST} astrometry, knowing that these  observations 
would also be able to
uncover any additional planetary companions with radial velocities greater 
than 4 \ms and
orbital periods of a year or less.  The HET High Resolution Spectrograph (HRS),
used in conjunction with an I$_2$ gas absorption cell, can give 
routine radial velocity precision of 3 \ms 
for high signal-to-noise observations \citep{Coc04}.  
By taking advantage of the 
queue-scheduled operation of the HET, we recorded well over 
100 radial velocity measurements in a span of
about 190 days, shown in Table \ref{tbl-rvdat}. This "snapshot" of the \rc system promised the best 
definition of the orbital parameters of the two known short-period 
inner planets, because it would be
relatively free of time-varying changes, not only in  the planetary 
system but also in the
facilities used to observe the system.

\begin{figure}
\epsscale{.80}
\plotone{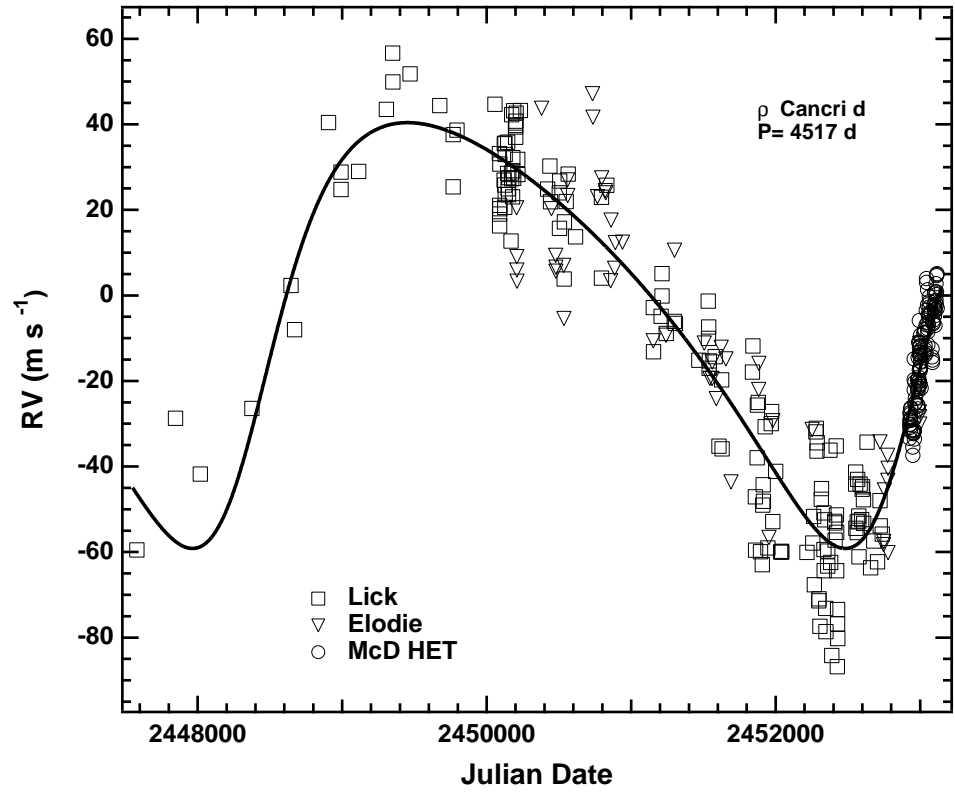}
\caption{The RV orbit of the long period planet (\rc d) is
shown with the three data sets.  \rc b and \rc c are
subtracted from the observations.  The rms of the HET residuals in
this fit are 9.2 \ms.
\label{fig-planetd}}
\end{figure}

The long period planet (\rc  d) could clearly not be modeled by the 
HET data alone, but because of its velocity amplitude it was a very 
critical part of  the initial modeling of the triple-Keplerian orbit.  
Three sets of radial velocity data from Lick \citep{Mar02},  
ELODIE \citep{Nae04} and
HET were fit with a triple-Keplerian orbit using Gaussfit \citep{Jef87}, 
employing the techniques to
combine RV from different sources \citep{Hat00} producing the orbit of the 
outer planet shown in Figure  \ref{fig-planetd}.   This orbital fit remains 
problematic because of some discontinuity between
the data sets for \rc c (P=44d), and the sparse early Lick data.
It is clear
that the extrema of the HET data are not fit as well  as the central portions. 
The orbit derived for this long period planet was then used as a 
constant in further modeling.

\begin{figure}
\epsscale{.80}
\plotone{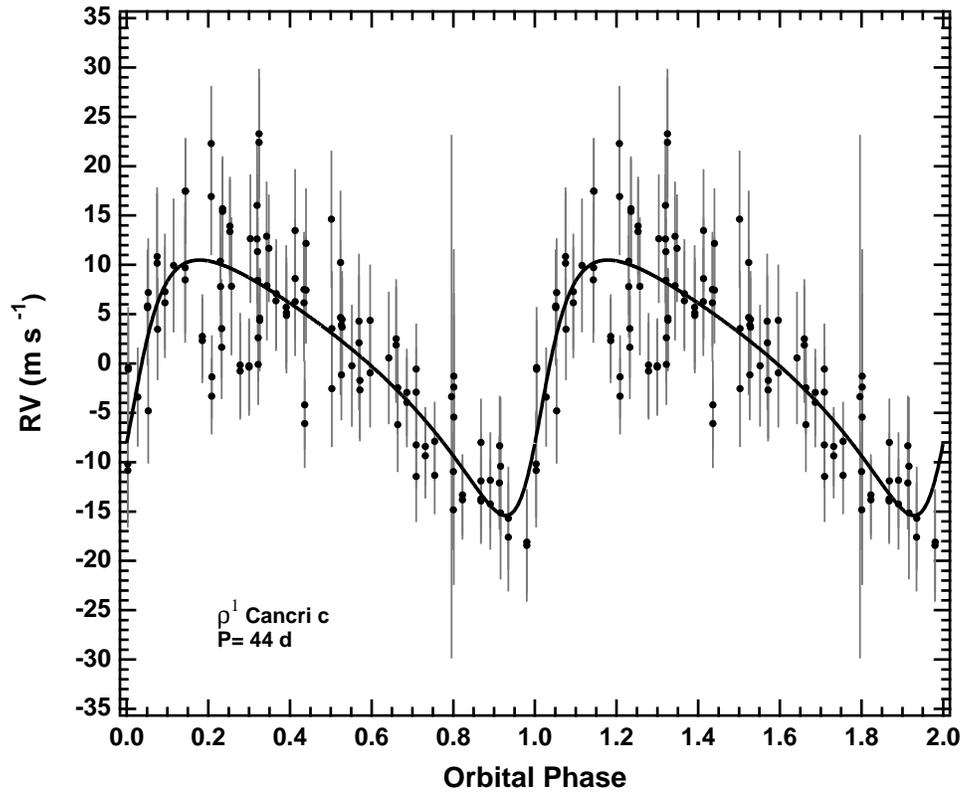}
\caption{The RV orbit of the $~44$ day (\rc d) planet 
is shown with the phased HET data.  Two cycles are plotted.
The signatures of \rc b, \rc d, and \rc 
have been subtracted from the observations.
\label{fig-planetc}}
\end{figure}

\begin{figure}
\epsscale{.80}
\plotone{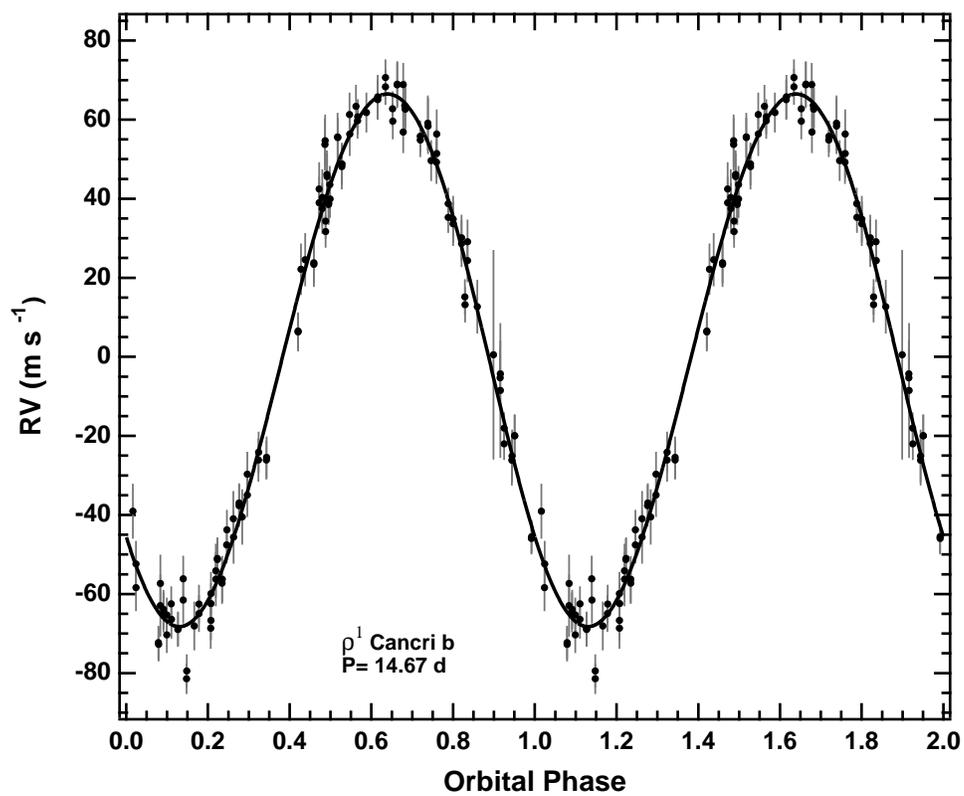}
\caption{The RV orbit of the $~14.66$ (\rc b) day  planet 
is shown with the phased HET data. 
\rc c,  \rc d,  and \rc e  are subtracted from the     observations.
\label{fig-planetb}}
\end{figure}

\rc c was discussed \citep{Mar02} as a 
signal that could possibly be caused by rotation modulation of 
inhomogeneous surface features.
The stellar rotation period of 36-42 days was uncovered from periodicities 
in the Ca II emission \citep{Hen00, Noy84}.  However, the planetary hypothesis is 
more likely because \rc  is a
very inactive star in which stellar activity induced RV effects generally do not
exceed 3.0 \ms. The amplitude \citep{Mar02} of the signal for this intermediate planet 
was over 12.0 \ms.  The
phased HET data showing \rc c are shown  in Figure \ref{fig-planetc}.  Figure  
\ref{fig-planetb} shows  \rc b, the 14.66 day period planet, the original
planet found around \rc.

\begin{figure}
\epsscale{.80}
\plotone{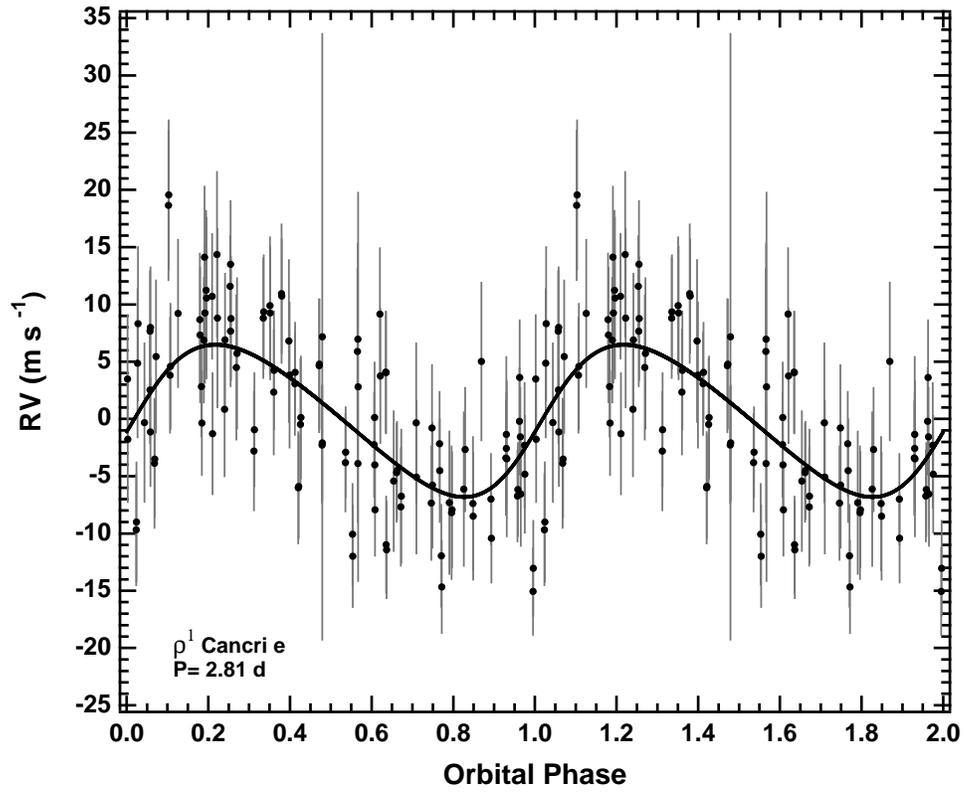}
\caption{The RV orbit of the 2.808 day  planet (\rc e)   is shown
with the phased HET data.  The \rc b, \rc d, and \rc d  planets are
subtracted from the    
observations.  The rms of the HET residuals in this fit are 5.4 \ms.
\label{fig-planete}}
\end{figure}

\begin{figure}
\plotone{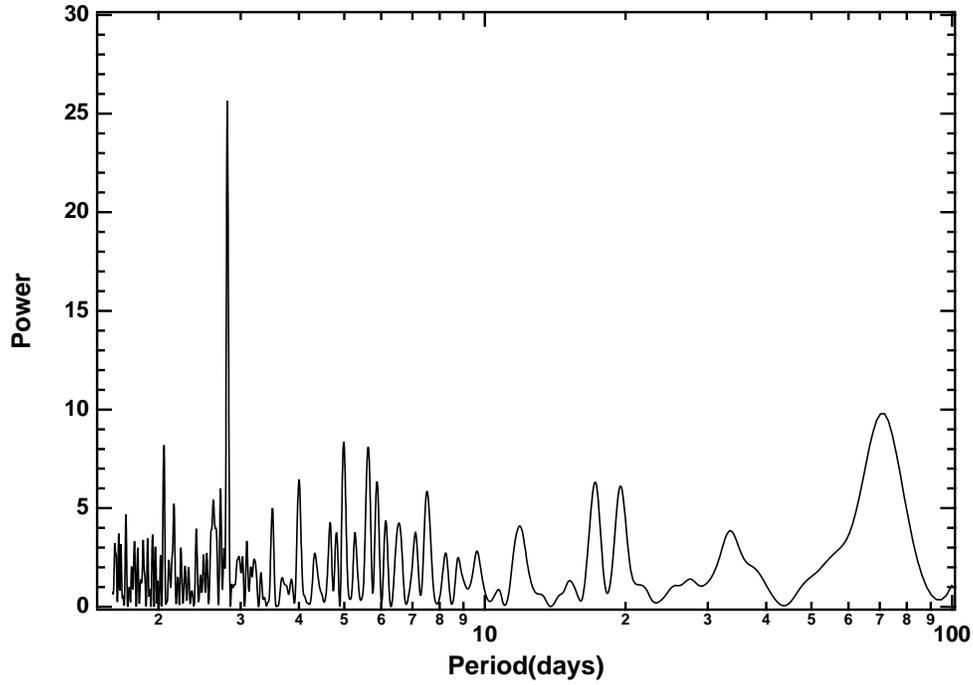}
\caption{Periodogram of the velocity residuals of the HET data, after the orbits of 
\rc b, \rc c, and \rc d have been subtracted from the original velocities. 
The peak at 2.808 days has a false alarm probability of  1.730796e-09.
\label{fig-powere}}
\end{figure}

\begin{figure}
\plotone{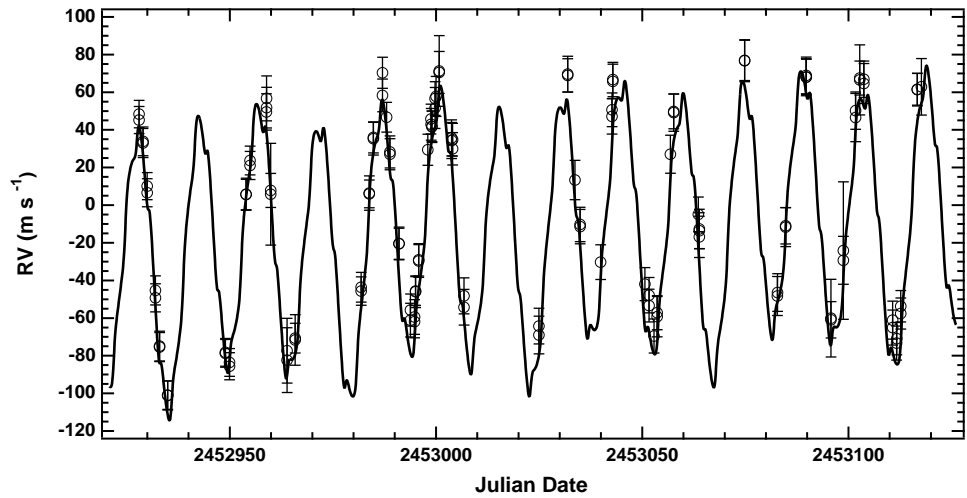}
\caption{The Keplerian RV orbit of the 4 planets (combined) around \rc   is 
shown
with the  HET data. 
\label{fig-allplanet}}
\end{figure}

Analysis of the residuals of the HET data, after removal of the best 
overall fit to the orbits of the three known  planets to all 
of the radial velocity and astrometric data,  revealed  a 
velocity perturbation with a period of 2.808  days.   
Simultaneous modeling of \rc b, \rc c, and \rc e with \rc d as a constant, 
showed 
an object in a low eccentricity orbit with a period of 2.808 $\pm$ 0.002 days. 
The radial velocity  semi-amplitude of this orbit is 
only 6.67  $\pm$ 0.81 \ms , which is above the 
expected internal radial velocity "jitter" variability for this star.  
The 2.808 day radial velocity signal is
coherent over the entire time-span of the HET data, which would not be expected
for stochastically excited  stellar surface sources.  The HET residuals, phased 
to the 2.808 day period with the orbital solution over plotted are show in 
Figure \ref{fig-planete}.  In retrospect, the clear
signal of this 2.808 day orbital period planet can be seen 
(along with the signal of the 44 day periodicity (\rc c)) in Figure 9 of 
Marcy \etal 2002, which shows the periodogram of the residuals
of the Lick data, after the \rc b and d have been removed.  This gives 
independent confirmation that this radial velocity periodicity is 
intrinsic to the \rc system, and is not an artifact of the observation or 
data analysis techniques used on either the HET or Lick data.  The 2.808 day 
periodic signal has a false-alarm probability of $1.7 \times 10^{-9}$
in the HET data (as seen in Figure \ref{fig-powere})  and 
$1.1 \times 10^{-6}$ in the Lick data, 
giving further strong evidence for its reality.
We interpret this 2.808 day radial velocity periodicity as the stellar reflex 
motion due to an extremely low mass planet in orbit around the star.  
Properties of the combined four-planet orbital solution to all of the radial 
velocity data are given in Table \ref{tbl-rvorb}. Mass parameters and limits are
shown in Table \ref{tbl-massprm}.  The HET data in time overplotted with the 
quad-Keplerian fit are shown in Figure \ref{fig-allplanet}.  The mass function 
from the radial velocity solution, combined with our
assumed stellar mass of 0.95 $\pm$ 0.08 solar masses, gives an \msini of this innermost 
2.808 day planet of 0.045 $\pm$ 0.01 \mjup.  If we assume that all of the planets in 
the \rc  system are coplanar and adopt the 53$\arcdeg$ inclination of the outermost 
planet, then we compute a true mass for this innermost planet of about 
17.7 Earth masses, close to the mass of Neptune. (If an inclination of 20$\arcdeg$
is assumed, an upper limit of 2.37 \mnep is derived.)
This fourth planet, \rc b, is the lowest mass extrasolar planet yet found
around a solar-type star.  We note that the lowest mass, and indeed the 
very first, extrasolar planets ever found are those around the 
millisecond pulsar PSR B1247+1221\citep{Wol92}.

\section{Discussion and Conclusion}
The  inclination of the \rc e, assuming coplanarity of the system ,
indicates that this object is indeed
a Neptune-mass object, and not a more massive planet viewed nearly face-on
to its orbit.   The presence of this very low-mass planet at such a small
semi-major axis in a system of at least three other gas-giant planets
provides an important testbed for models of planetary system formation
and evolution.

A crucial question is whether \rc e originally formed at something near the
current observed mass, or whether it formed as an intermediate or much more 
massive gas giant that has lost a significant amount of  mass during the 
evolution of the system.  

It is reasonable to assume that at least the three inner planets of
the \rc system were formed at roughly the same time, and that their dynamical
evolution was closely linked.  The most massive of these three planets,
\rc b (P=14.66d), would have been the first to achieve a mass 
sufficient to open a gap in the disk and begin Type~II migration 
\citep{War96}.  If this happened before the inner \rc e reached
the critical mass for rapid accumulation of a gas envelope,
then it could have been carried inward as part of the overall migration
of the disk material without ever growing past the rocky core phase
\citep{Lin86}.   Whether this inward tidal migration driven by \rc b
would sweep such a Neptune-mass planet into the stellar photosphere, or
would leave it in its present short period orbit depends on the details
of the end-process of planetary tidal evolution.
If the planet formed at a mass near its present mass by the core-accretion
model of planetary system formation \citep{Pol96},
it would be a rocky body which could have a substantial dense atmosphere.

An alternative hypothesis is that \rc e formed as a much more
massive gas giant planet before it began its inward type~II migration
to its present orbital radius \citep{Ida04}.  With the planet then
within 0.05 {\scshape au} of the parent star, it could have suffered
significant heating from tidal interactions with the star \citep{Bod01}.
This tidal heating could have caused the radius of the planet to inflate,
possibly overflowing the Roche radius of the star and resulting in
significant planetary mass loss \citep{Gu03,Bar04}.
It is possible that the mass loss has
stripped  this planet of almost all of its original H$_2$-He gas envelope,
leaving just the rocky planetary "core".  This loss of a gas envelope for
very short period planets is supported by the observation of an extended
H~I, O~I and C~II atmosphere around HD209458b (Vidal-Madjar {\itshape et~al.\/}
2003; 2004), which may be in hydrodynamic escape from the planet \citep{Lec04}.
However, the discovery of transiting Jupiter-mass planets (OGLE-TR-113 and
OGLE-TR-132) with orbital
periods of less than 2 days \citep{Bou04} may cast doubt on this scenario.
Both of the OGLE planets are about one Jupiter mass
and are only slightly greater than one Jupiter radius.  This indicates that
planets which are much more massive than \rc e can indeed survive
at even shorter semi-major axes.

We have shown that an intense campaign of directed observations
of a planetary system, aided by previous observations from other telescopes,
can uncover Neptune mass objects.  While it will be years before
astrometric missions (SIM and GAIA) will be able to  uncover the astrometric
perturbations orbits of all these planets, we have been able to
use {\it HST} data to place a preliminary limit on the inclination of the outer
planet.

\acknowledgments

The Hobby-Eberly Telescope (HET) is a joint project of the University of
Texas at Austin, the Pennsylvania State University,  Stanford University,
Ludwig Maximillians Universit\"{a}t M\"{u}nchen, and
Georg August Universit\"{a}t
G\"{o}ttingen.  The HET is named in honor of its principal benefactors,
William P. Hobby and Robert E. Eberly.
These results are partially based on data obtained with the UCO/Lick Observatory
Telescope,  the ELODIE spectrograph
at Observatoire de Haute-Provence (OHP, France), the Apache Point
Observatory 3.5m telescope which is owned and operated by the Astrophysical Research Consortium),  and the NASA/ESA 
Hubble Space Telecope (obtained  at the  the Space Telescope Science 
Institute, 
which is operated by  the Association of Universities for Research in 
Astronomy, Inc).
This paper is based upon  work supported by the NASA 
under grants  HST GO-09969, NAG5-13206 and NNG04G141G, and NSF 
under grant AST-9808980.
We are grateful to Ed Nelan for providing the {\it HST} data and Melissa
McGrath, for supporting our further investigation, and
to W. Spiesman for his careful reading of this manuscript.
We are grateful to our anonymous referee for their thoughtful comments.


%% copy editing.  Individual instruments can be provided in parentheses,
%% after the keyword, but they will not be verified.

Facilities: \facility{McD Obs(HET)}, \facility{HST(FGS)}, 
\facility{UCO/Lick)}, \facility{OHP(ELODIE)},  \facility{ARC(Apache Point)}.

\clearpage

\clearpage
\begin{deluxetable}{lcc}
\tablecaption{Measured Velocities for \rc \label{tbl-rvdat}}
\tablewidth{0pt}
\tablehead{
\colhead{Julian Date}    &
 \colhead{Radial Velocity} & \colhead{Uncertainty}\\
\colhead{} & \colhead{\ms} & \colhead {\ms}  }
\startdata
2452927.984100  &       48.3959 &       4.42    \\
2452927.986693  &       45.2119 &       4.61    \\
2452928.973951  &       32.9745 &       4.33    \\
2452928.976555  &       33.8595 &       4.00    \\
2452929.976107  &       10.1059 &       4.01    \\
\enddata
\tablecomments{[The complete version of this table is in the electronic edition of
the Journal.  The printed edition contains only a sample.]}

\end{deluxetable}

\clearpage
\begin{deluxetable}{lrrrr}
\rotate
\tablecaption{Quad-Keplerian Orbital  Elements of \rc \label{tbl-rvorb}}
\tablewidth{0pt}
\tablehead{
\colhead{Element}    &
 \colhead{\rc e } & \colhead{\rc b} & \colhead{\rc c} & \colhead{\rc d}  }
\startdata
Orbital Period {$\it P$} (days) &  2.808 $\pm$  0.002  & 14.67 $\pm$  0.01
  & 43.93 $\pm$ 0.25 & 4517.4 $\pm$ 77.8\\
Epoch of Periastron {$\it T$} \tablenotemark{a} & 3295.31 $\pm$ 0.32 &  
3021.08 $\pm$ 0.01 & 3028.63 $\pm$ 0.25 &  2837.69 $\pm$ 68.87 \\
Eccentricity  {$\it e$}& 0.174 $\pm$ 0.127 & 0.0197 $\pm$ 0.012 & 0.44 $\pm$
 0.08 & 0.327 $\pm$ 0.28 \\
$\omega$ ($\arcdeg$)& 261.65 $\pm$ 41.14 & 131.49 $\pm$ 33.27 & 244.39 $\pm$
10.65 & 234.73 $\pm$ 6.74  \\
Velocity amplitude {$\it K$} (\ms) & 6.665 $\pm$ 0.81  & 67.365  $\pm$ 0.82  &
12.946 $\pm$ 0.86 & 49.786 $\pm$ 1.53 \\
$V_0$ Lick (\ms) &  21.166 $\pm$ 1.31 \\
$V_0$ ELODIE (\ms) & 2727.448 $\pm$ 2.42 \\
$V_0$ HET (\ms) & 10.745 $\pm$ 0.59
\enddata
\tablenotetext{a}{Add 2450000.0 to T}

\end{deluxetable}

\clearpage

\begin{center}
\begin{deluxetable}{lrrrr}
\rotate
\tablecaption{\rc  - Mass Limits and Parameters \label{tbl-massprm} }
\tablewidth{0pt}
\tablehead{
\colhead{Parameter}    &
 \colhead{\rc e} & \colhead{\rc  b} & \colhead{\rc c} & \colhead{\rc d}  }
\startdata
{\it a} (AU) & 0.038 $\pm$ 0.001 & 0.115 $\pm$ 0.003 & 0.240 $\pm$ 
0.008 & 5.257 $\pm$ 0.208\\
\asini (AU) & 1.694e-6 $\pm$ 0.19e-6 & 9.080e-5 $\pm$ 0.12e-5 & 4 .695e-5 $\pm$ 0.14e-5 & 0.195e-1 $\pm$ 0.007e-1\\
Mass Fraction (\msun)& 8.225e-14 $\pm$ 2.33e-14 & 4.64e-10 $\pm$ 0.17e-10
& 7.151e-12 $\pm$ 0.54e-12 & 4.874e-08 $\pm$ 0.38e-8\\
\msini (\mjup)\tablenotemark{a}  &  0.045 $\pm$ 0.01 & 0.784 $\pm$ 0.09 &  
0.217 $\pm$ 0.04  &   3.912 $\pm$ 0.52\\
\msini (\mnep)\tablenotemark{a} &  0.824 $\pm$ 0.17& & &\\
\msini (\mearth)\tablenotemark{a}&   14.210 $\pm$ 2.95& & &\\
M (\mjup)\tablenotemark{b,d} & 0.056 $\pm$ 0.017 & 0.982 $\pm$ 0.19 & 
0.272 $\pm$  0.07 & 4.9 $\pm$ 1.1\\
M (\mjup)\tablenotemark{c,d} & 0.053 $\pm$ 0.020 & 0.982 $\pm$ 0.26 & 0.244 
$\pm$ 0.07 & 4.64 $\pm$ 1.3\\
M  (\mnep)\tablenotemark{c,d}   &  1.031 $\pm$ 0.34& & &\\
M  (\mearth)\tablenotemark{c,d}  &  17.770 $\pm$ 5.57& & &\\
\enddata
\tablenotetext{a}{derived from radial velocity alone}
\tablenotetext{b}{derived from radial velocity and astrometry, using Msini/sini}
\tablenotetext{c}{derived from radial velocity and astrometry, using 
$m2^3/(m1 + m2)^2 = a^3/P^2$}
\tablenotetext{d}{assumes coplanarity of the planetary system}
\end{deluxetable}
\end{center}

\clearpage

\clearpage

\clearpage

\clearpage

%% The following command ends your manuscript. LaTeX will ignore any text

%temark{b,d}

\end{document}